\documentclass{aastex631}

\begin{document}

\title{Weighted Stacking of Radio Images Affected by Noise and Interfering Signals}

\author[0009-0002-0149-9328]{Giulia Murgia}
\affiliation{Division of Physics, Mathematics and Astronomy\\ California Institute of Technology \\ Pasadena, CA 91125, USA}

\author{Sofia Fatigoni}
\affiliation{Division of Physics, Mathematics and Astronomy\\ California Institute of Technology \\ Pasadena, CA 91125, USA}

\begin{abstract}
We implement an algorithm based on the weighted stacking of astronomical images that can combine different observations of the same region of the sky removing the interfering signals. We develop a C++ code that takes as input a set of spectral cubes and computes the local weights of the intensity for each pixel of every channel. The weights are calculated as the inverse variance of the nearby pixels and are used to compute the weighted merge of the input files. Astronomical sources, present in all cubes, are preserved by the weighted average. However, interfering signals, present in specific cubes and a certain frequency range, are down-weighted in the average and removed from the output spectral cube. We present the results obtained by analyzing simulated spectral cubes containing astronomical sources, noise, and a random set of interferences of different intensities and spectral occupations. The performance of the algorithm is evaluated by comparing the output result with the input sky model image. Finally, we present the results obtained by applying the method to a set of real data consisting of observations of the Andromeda galaxy, Messier 31 (M31), at 6.6 GHz obtained with the Sardinia Radio Telescope.
\end{abstract}


\section{Introduction} \label{sec:intro}

Radio Frequency Interference (RFI) presents a significant challenge to modern astronomical observations, particularly in the radio frequency spectrum. The increasing prevalence of terrestrial services and the rise of satellite constellations for radio communications have intensified this issue \citep{Lang23}. These man-made interferences have the potential to distort or obscure the faint cosmic emissions that astronomers aim to detect. Efficient Radio Frequency Interference (RFI) detection and removal is crucial to ensure the accuracy and reliability of astronomical data.

RFI comprises extraneous radio signals emanating from human-made sources, such as communication devices, radar systems, broadcast transmissions, and even electronic appliances. One significant source of RFI is the widespread use of wireless communication devices. Mobile phones, Wi-Fi routers, and other similar devices operate by transmitting and receiving radio frequency signals. Additionally, satellite communication and radar systems contribute to RFI \citep{Hainaut20}. These systems employ powerful transmitters that emit strong radio frequency signals. While essential for their respective applications, they can inadvertently reach the sensitive instruments of radio telescopes and interfere with the observations.

The spectrum of RFI encompasses a wide range of frequencies, spanning from the lowest bands used for power distribution to the microwave frequencies harnessed for telecommunications. The propagation of RFI through the Earth's atmosphere is a complex interplay of physical phenomena. Factors such as frequency, atmospheric conditions, and local topography influence the behavior of these signals. They can refract, diffract, and scatter, producing intricate patterns of interference. Signals can arrive at radio telescopes from multiple directions, often with varying intensities and polarizations, further complicating their detection and mitigation.
 
The spurious signals from RFI introduce a layer of noise that alters the astronomical emissions of interest. Therefore, RFI removal is an important step in the data analysis pipelines of radio observations. 
Traditional approaches to mitigating RFI have encompassed various techniques \citep{Fridman01}. Bandpass filters, for example, are used to selectively reduce specific frequency ranges that are susceptible to interference. However, their application requires careful consideration, as they can unintentionally reduce genuine astrophysical signals. Additionally, time-domain filters and spectral excision algorithms have been used, though they can introduce artifacts or produce false positives \citep{grobler2016, Chung13, Khaliel21}.

Adaptive filtering techniques have also been explored. They take advantage of the dynamic nature of RFI, utilizing its distinct characteristics to differentiate it from astrophysical signals.  Machine learning algorithms, trained on large datasets of RFI examples, have shown promise in autonomously identifying and removing interference patterns. However, the effectiveness of these methods depends on having sufficient training data and the ability of the algorithm to adapt to evolving interference patterns \citep{mitchellwynne2018, Mosiane17, Harrison19}.

Another approach involves using redundant baselines in interferometric observations. By comparing the signals from multiple antennas, it is possible to identify and isolate spurious signals that do not align with the expected correlation patterns of celestial sources \citep{Offringa10}.

Despite these advances, no one-size-fits-all solution to RFI mitigation exists, and each approach carries its strengths and limitations. Therefore, the development of innovative methodologies, such as the algorithm presented in this study, remains critical to advancing the field of radio astronomy.

\section{Algorithm and Program Code} \label{sec:Algorithm}

We implement a new algorithm based on the weighted stacking of astronomical images that is able to combine different observations of the same region of the sky removing the interfering signals.
We develop a C++ code that takes as input two or more spectral cubes and computes the local weights of the radio intensity for each pixel of every channel. The weights are computed as the inverse variance of the nearby pixels. An output spectral cube is obtained by the weighted merge of the input files. Real sources, which are present in all cubes, are preserved by the weighted average. However, interfering signals, which are present in specific cubes and a certain frequency range, are down-weighted in the average and are removed from the output spectral cube.

\subsection{The Spectral Weighting Algorithm}
\label{Algorithm}

The basic idea behind the algorithm is based on the knowledge of the properties of the noise in every channel of the spectral cubes. The spectral cubes are a stack of images each corresponding to a different spectral channel.
Spectral cubes have two positional dimensions, $x$ and $y$, and a third spectral dimension, $c$. The two first axes are usually the Right Ascension (RA), and Declination (decl.), while the third axis is the frequency or the spectral channel number. The pixel values refer to the sky brightness at the given position and frequency. 
For the sake of simplicity, in the following, we refer to "images" as the images of the same channel $c$ in different spectral cubes. Without loss of generality, we may assume that in regions of the sky with no RFI signals nor real sources, the distribution of pixel intensity of the noise is expected to be a Gaussian with standard deviation $\sigma_N$.
On the other hand, close to those pixels affected by RFI signals or real sources, the statistics of the pixel values deviate from the one expected for noise distribution. In particular, the local dispersion is expected to be larger than $\sigma_N$. 
This could be used to remove the RFI if two or more observations of the same region of the sky are available.
Let's assume that real sources are present in all the images (i.e. they are constant in the considered time interval), while the RFI signals are present just in one of them. If this is the case, we can attempt to combine the images using a local weighted average of the pixel values. 
For every pixel  $(x,y,c)$ in the images, we gather surrounding pixel values within a square region and calculate the local standard deviation, denoted as $\sigma_{L}(x,y,c)$. The square region's size is determined as $1+2L$ pixels, where $L$ is a user-selectable parameter. This choice ensures an odd number of pixels per side, maintaining symmetry around the central pixel. The total number of pixels in the region is $(1+2L)^2$.

From this sigma image, we determine the inverse-variance weights image as $w(x,y,c)=1/\sigma^2_{L}(x,y,c)$. The weight is low where $\sigma_{L}(x,y,c) \gg \sigma_N$, which means in regions near RFIs or real sources. 
Instead, the weight is high in the noise, where $\sigma_{L}(x,y,c) \simeq \sigma_N$. 
Note that the actual value of the local mean of the noise distribution is not relevant for the calculation of the weights. The crucial factor is the dispersion of the pixel values relative to the local average, rather than the local average itself.

Finally, an output image for channel $c$ is obtained by computing the weighted average of the input images taken at different times.  
The weighted average is computed as 
\begin{equation}
    avg(x,y,c)=\frac{\sum\limits_{i=1}^{N} v_{i}(x,y,c)\cdot w_{i}(x,y,c)}{\sum\limits_{i=1}^{N} w_{i}(x,y,c)},  
    \label{Eq:WeightedAverage}
\end{equation}

where $N$ is the number of images, and $v_{i}(x,y,c)$ is the pixel value of image $i$ of spectral channel $c$. 
Since RFI have low weights and are present just in specific cubes and a certain frequency range, they are down-weighted in the average and are removed from the output spectral cube. Real sources and spectral lines also have low weights. However, they are present in every spectral cube, and therefore they are not penalized by the weighted average.

\subsection{Program Code}

We developed the C++ code-named \textsc{Spectral-Weighting} \citep{Murgia_Spectral-Weighting_24}. \textsc{Spectral-Weighting} can process spectral cubes in FITS format \citep{Fits}. The code has two main functionalities: simulations and spectral merging. 
The simulation mode is used to test the reliability of the merging algorithm. The code can create spectral cubes containing simulated sources, noise, and a random set of interferences of different intensities and spectral occupations.
The spectral merging accepts as input a list of FITS cubes, each representing observations of the same sky region taken at different times. It provides as output a single cube representing the weighted average of the input cubes. 
The user interacts with the code either with a command line or using a script (the "pipeline") where it is possible to set the parameters for the different functions and call the methods in a logical sequence.

\section{Simulations} \label{sec:Simulations}

\begin{figure}[b]
\begin{center}
\includegraphics[height=7cm, trim=70 70 70 70]{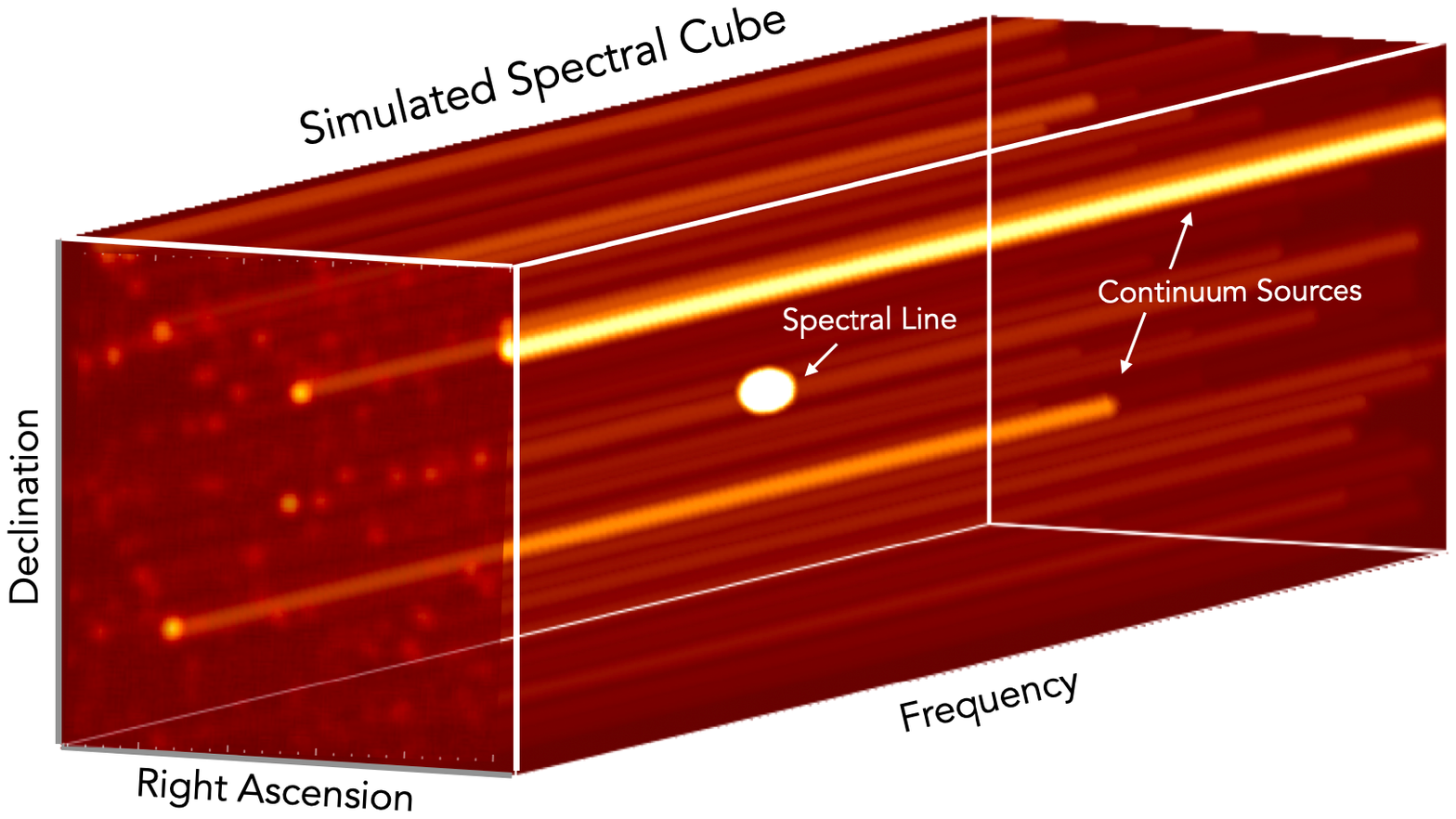}
\caption{Three-dimensional rendering of the spectral cube representing the simulated sky model. In the cube, continuum sources stretch along the frequency axis. The spectral line source is concentrated in a narrow range of channels at the center of the cube. 
\label{fig:CubeNoRFI}}
\end{center}
\end{figure}

\begin{figure}[t]
\begin{center}
\includegraphics[height=7.5cm]{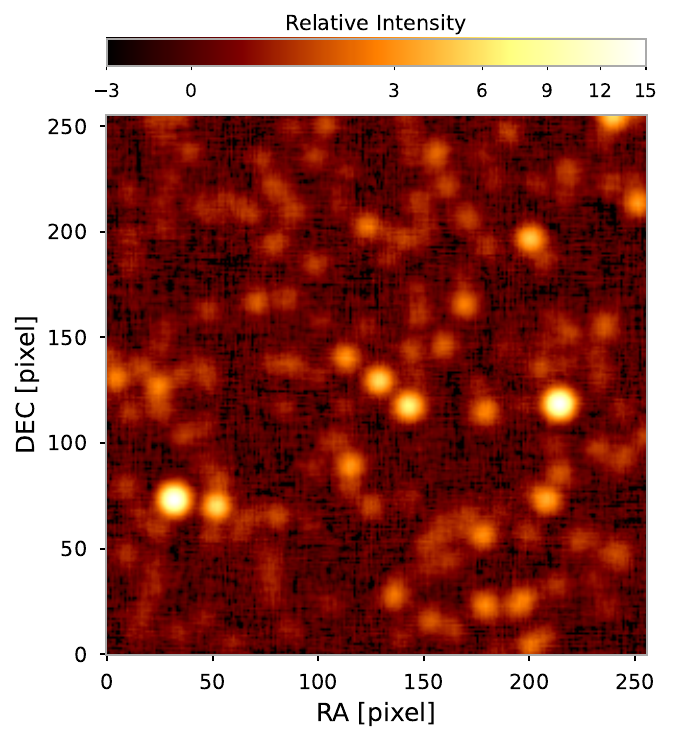}
\caption{Simulated sky model image with noise. The spectral average of a spectral cube was obtained by merging a simulated Right Ascension scan and a Declination scan. The simulated R.A. and decl. scans include a Gaussian noise and a spectral line at the center of the cube. The point sources are smoothed with a Gaussian kernel with $\sigma$ of 4 pixels. The color bars represent the intensity in arbitrary units relatively to 1-$\sigma$ noise level.
\label{fig:SplatMergeNoRFI}}
\end{center}
\end{figure}

\begin{figure}[b]
\begin{center}
\includegraphics[height=7cm, trim=70 70 70 70]{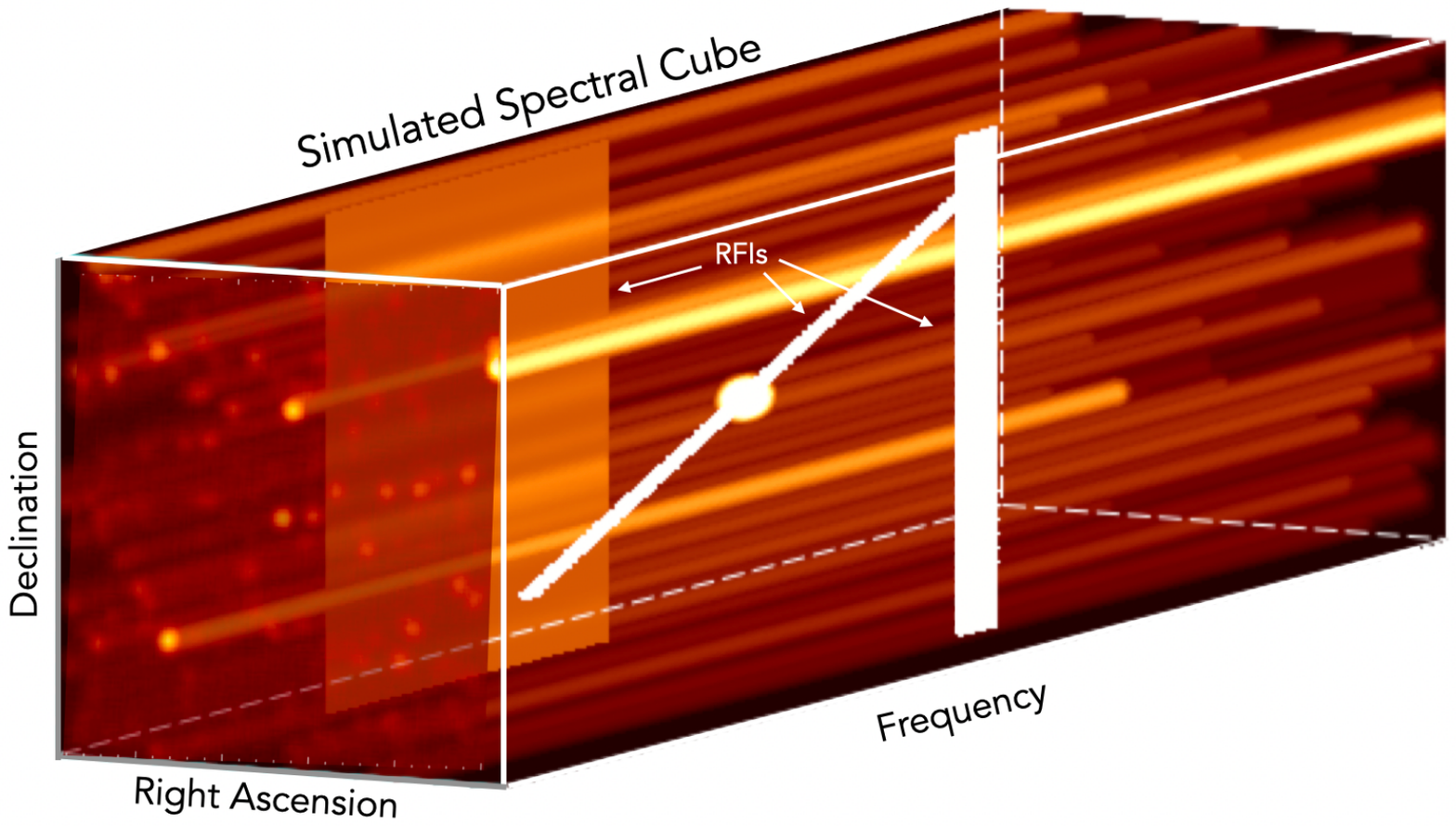}
\caption{Example of a three-dimensional rendering of the spectral cube representing the simulated sky model with RFIs. In the cube, RFI appears like planes or stripes.  
\label{fig:CubeRFI}}
\end{center}
\end{figure}

To verify the correct functioning of the algorithm, we apply it to simulated spectral cubes. The simulated spectral cubes contain astronomical sources, noise, and a random set of interferences of different intensities and spectral occupations. The performance of the algorithm is evaluated by comparing the output result with the input sky model image.

\subsection{Sky Model and Simulated RFIs}

First, we realize the simulated sky model that consists of a set of point-like sources with continuum spectra. These would represent the emission of individual galaxies and quasars that make up the background of the radio sky. 
For each source, we designate a random position for the R.A. and decl. coordinates. Then, we choose a random intensity according to the logN-logS distribution, simply assuming that $N(>S) \propto S^{-3/2}$. For all these continuum sources, we assume a spectrum $S_{\nu}=S_{0}\nu^{-\alpha}$, with $\alpha=0.8$, and we use a bandwidth of 10\% the frequency. This implies that the flux density of these sources gradually decreases by 7\%, spanning from the initial to the last channel of the cube. In addition, we introduce just one single source at the center of the image whose spectrum consists of a spectral line represented by a Gaussian profile $S_{\nu}=S_{0}e^{-\frac{1}{2}((\nu-\nu_{0})/\delta_{\nu})^2}$. We add a narrow Gaussian profile centered at $\nu_{0}$ corresponding to channel 500 and with a $\delta_{\nu}=8$ channels. The position of the spectral line source is at the center of the simulated image. We approximate the telescope beam pattern using a two-dimensional Gaussian profile with $\sigma = 4$ pixels. We obtain a smoothed version of the sky model by convolving each spectral channel image of the cube with the Gaussian beam. A three-dimensional rendering of the spectral cube representing the simulated sky model is shown in Fig. \ref{fig:CubeNoRFI}.

Next, we generate noise for the R.A. and a noise for the decl. cubes. This is because we want to simulate the radio telescope scans that are along the R.A. and decl. directions. Thus, for the R.A. direction (x-axis), we implement a correlated noise model of the form 
\begin{equation}
    noise(x_{i})=\eta \cdot noise(x_{i-1})+r
    \label{Eq:CorrelatedNoiseX}
\end{equation} 
where $r$ is a random value extracted from a Gaussian distribution with standard deviation $\sigma_{N}$, and $\eta$ is a fixed factor between 0 and 1. If $\eta=0$, the noise in each pixel is independent and there is no correlation. In this simulation set, we use $\eta=0.8$. Similarly, for the decl. direction (y-axis), we implement the same model as in Eq. \ref{Eq:CorrelatedNoiseX}.

The simulated sky model image is shown in Fig. \ref{fig:SplatMergeNoRFI}. It consists of the spectral average of a spectral cube obtained by merging a simulated Right Ascension scan and a Declination scan.

\begin{figure} [t]
\begin{center}
\begin{tabular}{c}
\includegraphics[height=7.5cm]{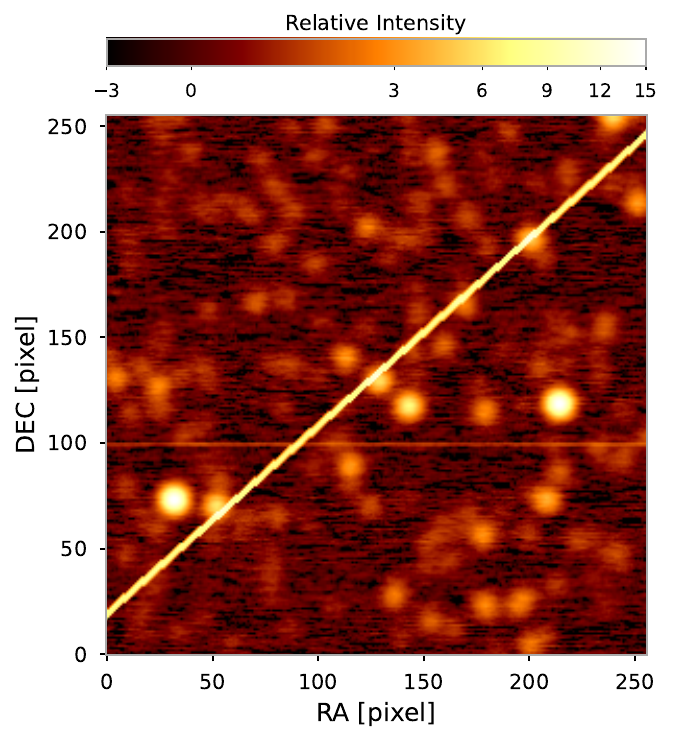}
\includegraphics[height=7.5cm]{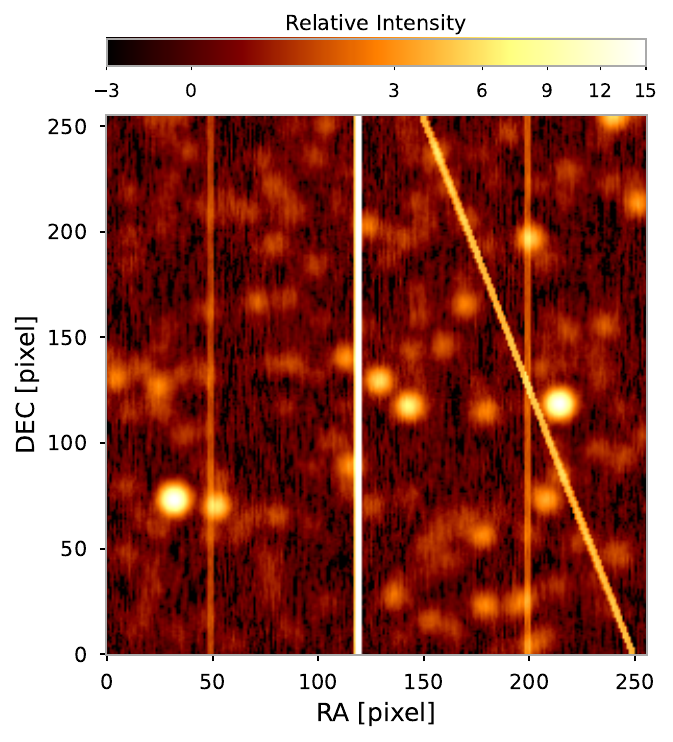}
\end{tabular}
\end{center}
\caption 
{ \label{fig:SplatCubes}
Spectral average of the simulated cubes that include RFIs. Left Panel: spectral average of the simulated Right Ascension scan. Right Panel: spectral average of the simulated Declination scan. The color bars represent the intensity in arbitrary units relatively to 1-$\sigma$ noise level.} 
\end{figure}

\begin{figure} [h!]
\begin{center}
\begin{tabular}{c}
% trim=left bottom right top
\includegraphics[height=7.5cm, clip]{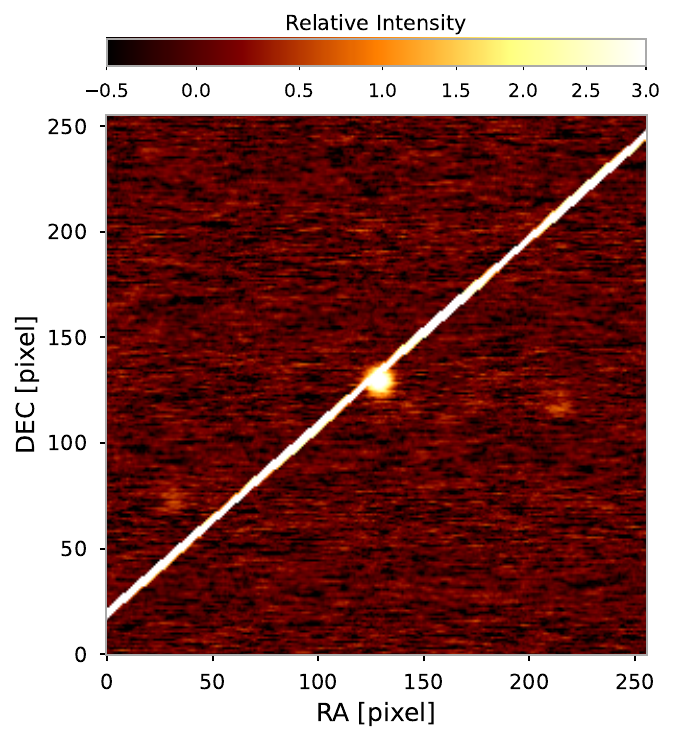}
\includegraphics[height=7.5cm, clip]{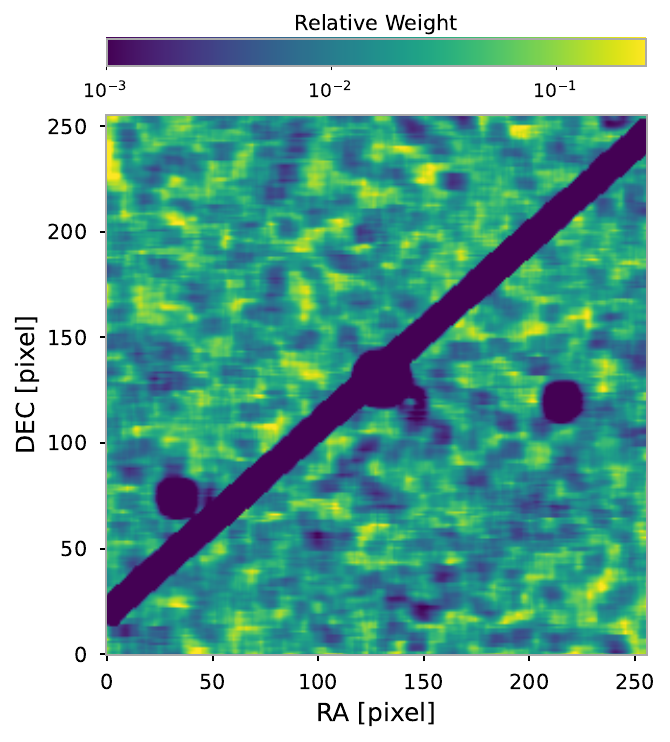}
\end{tabular}
\end{center}
\begin{center}
\begin{tabular}{c}
% trim=left bottom right top
\includegraphics[height=7.5cm, clip]{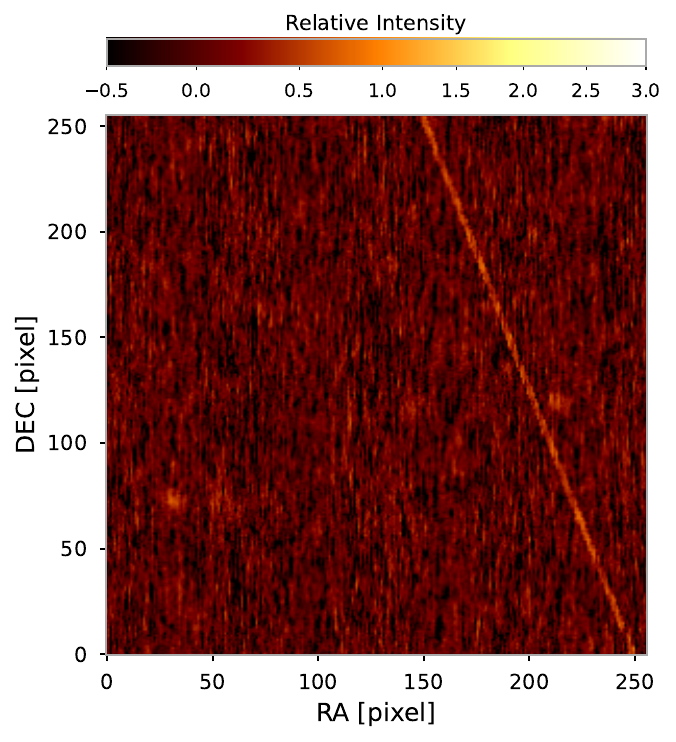}
\includegraphics[height=7.5cm, clip]{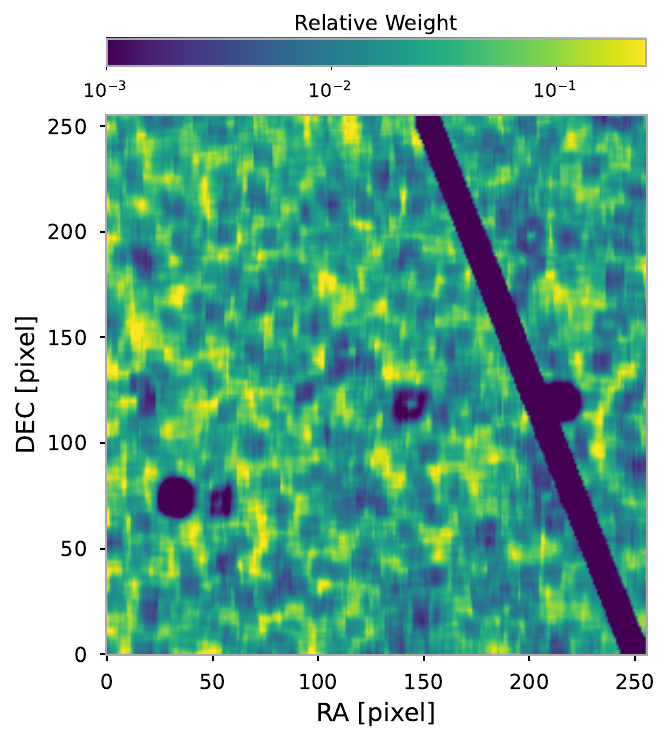}
\end{tabular}
\end{center}
\caption 
{ \label{fig:SimulationExample}
Sample channels from the simulated cubes. Top Panels: the R.A. scan for channel 510 is shown in the left panel; the image of the respective weights is shown in the right panel.
Bottom Panels: the decl. scan for channel 950 is shown in the left panel; the image of the respective weights is shown in the right panel.} 
\end{figure}

Finally, we introduce a set of simulated RFIs. They are defined by specifying the starting and ending positions, the channel range, and the intensity. The signals affect all the pixels along the line connecting the starting and ending positions. 
We simulate two different types of RFIs: those originated by terrestrial sources that last in time just for one telescope scan and thus are oriented along R.A. or decl., and those originated by satellites that could be inclined due to the motion of the interfering source. In addition, we simulate both RFIs that are extended in a broad channel range and that appear like planes in the three-dimensional representation, and RFIs that are narrow in frequency and appear like stripes. An example of a three-dimensional rendering of the spectral cube representing the simulated sky model with RFIs is shown in Fig. \ref{fig:CubeRFI}. An example of R.A. and decl. simulated images obtained by averaging spectral cubes is shown in Fig. \ref{fig:SplatCubes}. Each scan contains three RFIs with different orientations and intensities. 

\subsection{Testing the Spectral-Weighting Algorithm on Simulated Data}
\label{sec:TestingOnSimulations}

We run the \textsc{Spectral-Weighting} code on the simulated R.A. and decl. cubes. For each input spectral cube, the program computes the cubes of the local weights. Examples of the local weights for specific channels are shown in Fig. \ref{fig:SimulationExample}.

\begin{figure}[h!]
\begin{center}
\begin{tabular}{c}
\includegraphics[height=7.5cm]{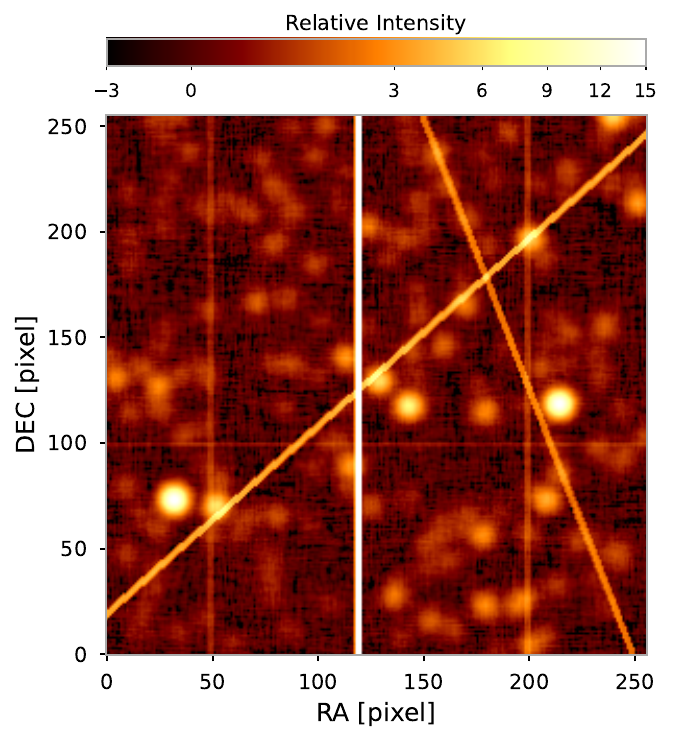}
\includegraphics[height=7.5cm]{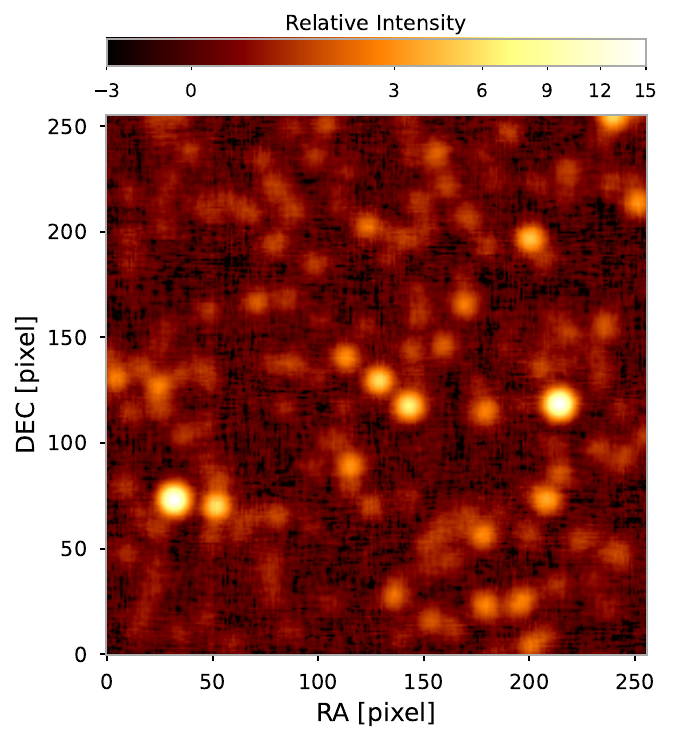}
\end{tabular}
\end{center}
\caption 
{ \label{fig:SimultionComparison}
 Images of the spectral average of the simulated cubes. Left panel: direct spectral average without removing the RFIs. Right panel: spectral average obtained by applying the weighted stacking algorithm presented in this work. The color bars represent the intensity in arbitrary units relatively to 1-$\sigma$ noise level.}
\end{figure}

Then, for each channel, the code computes the weighted average of the two scans and writes the merged image into the corresponding channel of the output spectral cube. The result is presented in the right panel of Fig. \ref{fig:SimultionComparison}, where we show the spectral average obtained by collapsing all the frequency channels in one image. For reference, in the left panel of Fig. \ref{fig:SimultionComparison}, we show the spectral average of the merge cube obtained without applying the algorithm. Comparing these two images, it is readily visible that the RFIs are efficiently removed. Finally, comparing the final output shown in the right panel of Fig. \ref{fig:SimultionComparison} with the original sky model image of Fig. \ref{fig:SplatMergeNoRFI}, we conclude that the algorithm works remarkably well in removing the RFIs. 

\begin{figure}[h!]
\begin{center}
\includegraphics[height=7.5cm]{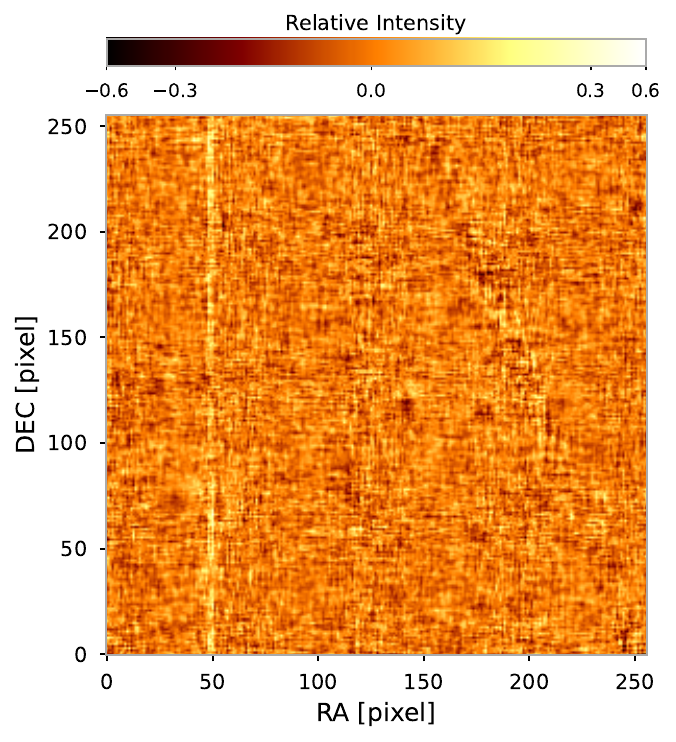}
\caption{Residuals obtained subtracting the original sky model image without noise and RFIs from the final output. 
\label{fig:Residuals}}
\end{center}
\end{figure}

The residuals obtained subtracting the original sky model image without noise and RFIs from the final output shown in the right panel of Fig. \ref{fig:SimultionComparison} is shown in Fig. \ref{fig:Residuals}. Overall, the majority of the RFIs are successfully removed. Indeed, the image of the residuals shows only noise, except for a very weak remaining vertical signal located at $RA=50$ pixel. It corresponds to a faint and broadband RFI that extends from channel 200 to channel 500. The original RFI is comparable to the noise of the individual channels, and therefore it is hard for the algorithm to completely down-weight it. Nevertheless, the residual signal from this RFI is only 0.57$\sigma_{N}$, where $\sigma_{N}$ is the noise standard deviation. 

\begin{figure}[h!]
\begin{center}
\begin{tabular}{c}
\includegraphics[height=5.5cm, clip]{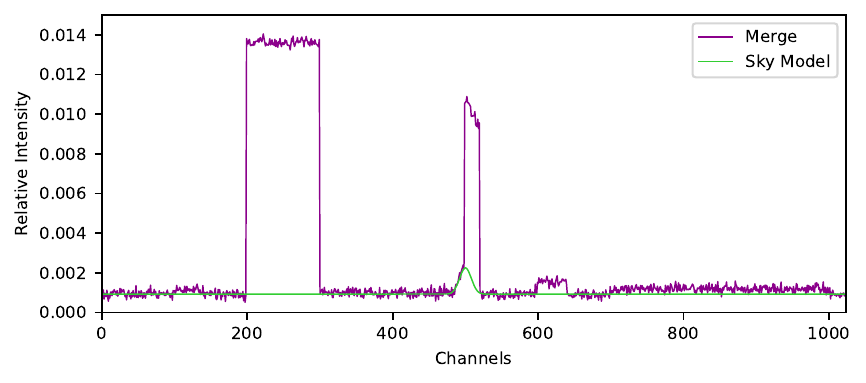}
\end{tabular}
\end{center}
\begin{center}
\begin{tabular}{c}
\includegraphics[height=5.5cm, clip]{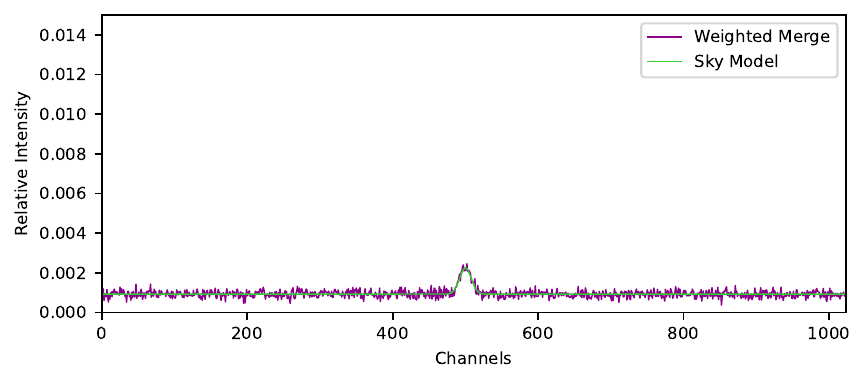}
\end{tabular}
\end{center}
\caption 
{ \label{fig:SimulationSpectra}
Simulated cubes spectra extracted from the spectral average images shown in Fig.\,\ref{fig:SimultionComparison}. Top Panel: spectrum of the direct spectral average without removing the RFIs. Bottom Panel: spectrum of the spectral average obtained by applying the weighted stacking algorithm. The green line represents the sky model, i.e. the signal from the simulated astronomical sources before adding the noise and RFIs.} 
\end{figure}

The spectra extracted from the merge cubes of Fig.\,\ref{fig:SimultionComparison} are shown in Fig. \ref{fig:SimulationSpectra}. In the top panel, we observe the effect of the  RFIs. All the RFIs have been removed in the bottom panel, but the spectral line source, which in the top panel was hidden from one RFI, is completely recovered. From these spectra, we conclude that the algorithm successfully removes all the RFIs preserving the spectral line emission.

\section{Application to real data} \label{sec:RealData}

We present the results obtained by applying the algorithm to a set of real data consisting of observations of the Andromeda galaxy, Messier 31 (M31), at 6.6 GHz obtained with the Sardinia Radio Telescope in 2016. The Sardinia Radio Telescope (SRT) is a single-dish radio telescope run by the Istituto Nazionale di Astrofisica, INAF. The observations of Messier 31 (M31), at 6.6 GHz, were completed in 64 hours, divided into 6 days, during which 22 complete scans of the galaxy were obtained.

\subsection{Pre-analysis of the data}

We apply the algorithm to calibrated data. For data analysis details, we refer to the work carried out by \cite{Fatigoni21} and \cite{Battistelli19}. Regarding the application presented in this work, it is relevant to point out that data are calibrated for bandpass and flux density. Moreover, a baseline subtraction has been applied to remove the receiver noise, the atmosphere emission, and the large-scale foreground and background emission from the Milky Way and the CMB radiation. 

In the works by \cite{Fatigoni21} and \cite{Battistelli19}, RFI signals were removed with automated flagging methods. Overall, approximately $30\%$ of the total data was removed due to RFIs.
However, for the purpose of this work, we do not apply those flags to the data because we want to test the \textsc{Spectral-Weighting} algorithm.

\begin{figure} [h!]
\begin{center}
\begin{tabular}{c}
% trim=left bottom right top
\includegraphics[height=7cm, clip]{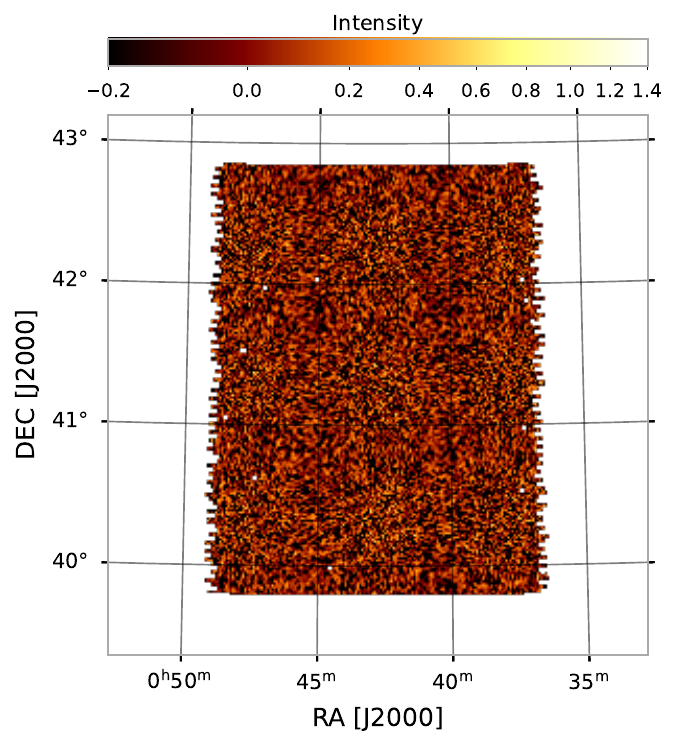}
\includegraphics[height=7cm, clip]{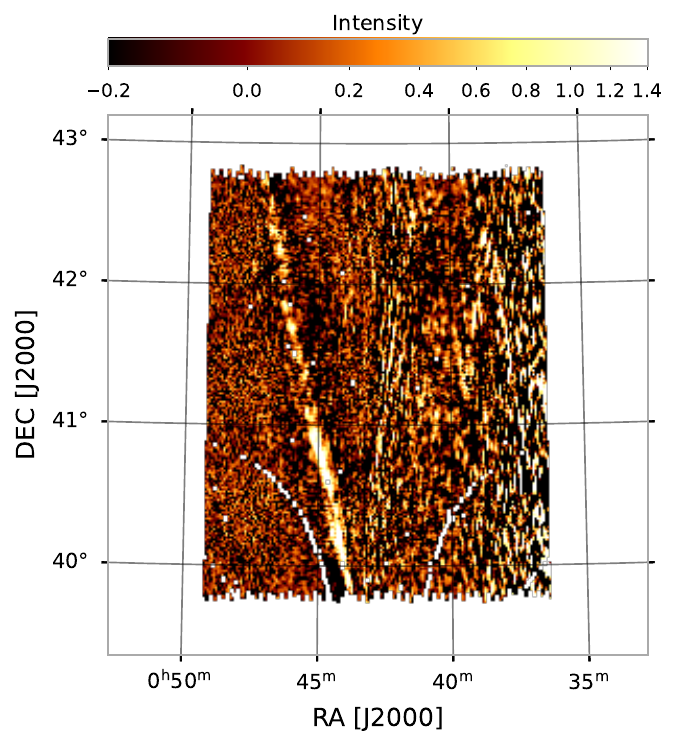}
\end{tabular}
\end{center}
\begin{center}
\begin{tabular}{c}
% trim=left bottom right top
\includegraphics[height=7cm, clip]{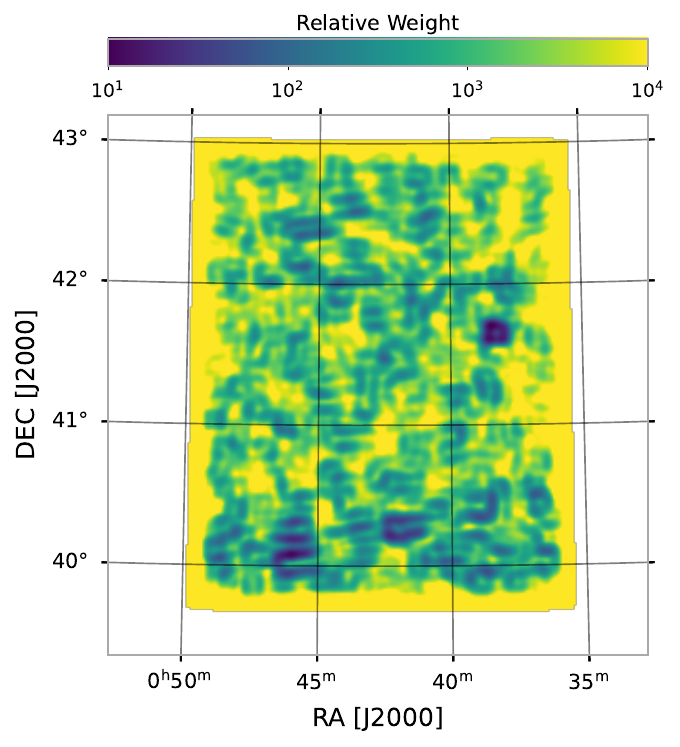}
\includegraphics[height=7cm,  clip]{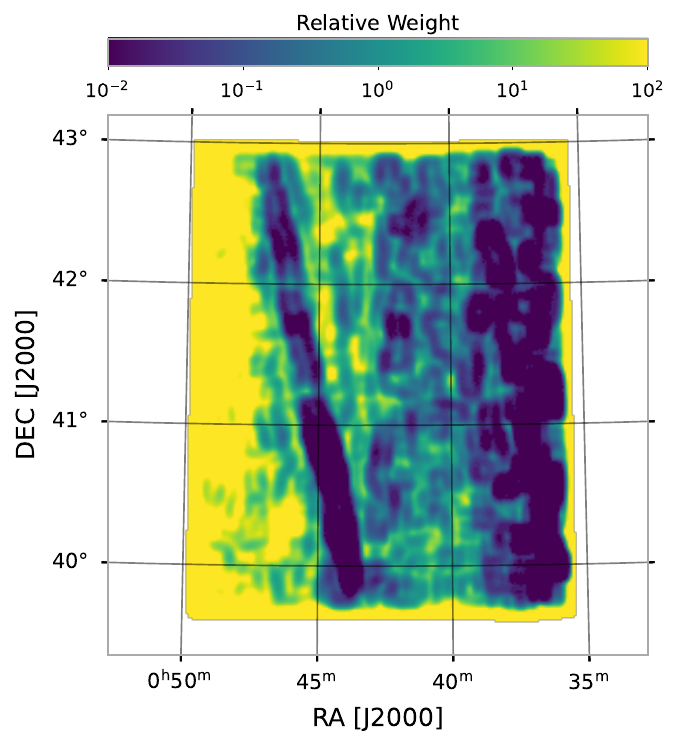}
\end{tabular}
\end{center}
\begin{center}
\begin{tabular}{c}
% trim=left bottom right top
\includegraphics[height=7cm, clip]{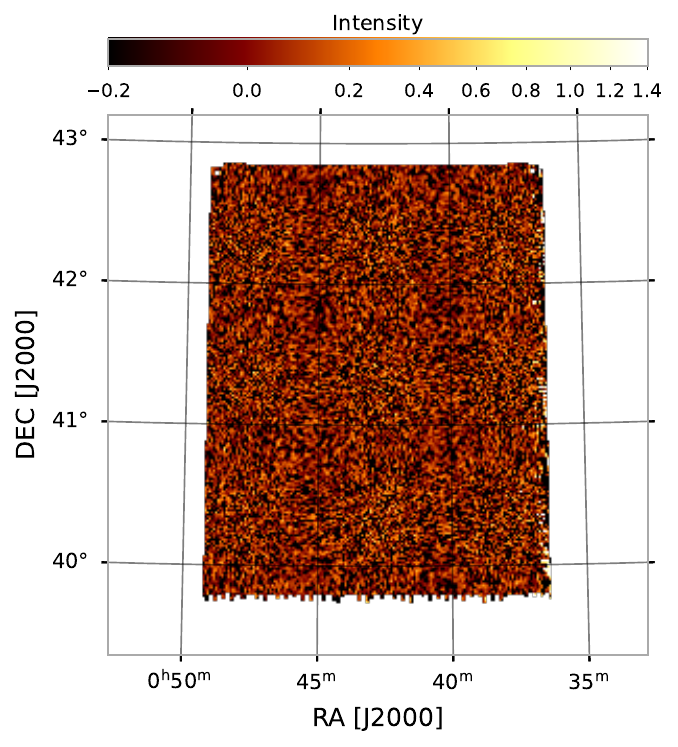}
\end{tabular}
\end{center}
\caption 
{ \label{fig:Channel2342}
M31 images of the spectral channel 2342 ($\nu=6.11$\,GHz). Top Panels: Right Ascension and Declination scans are shown in the left and right images. The color bars represent the intensity in JY/BEAM. Middle Panels: images of the weights used in the stacking for the R.A. and decl. scans. Bottom Panel: the image of the weighted merge of the R.A. and decl. scans.} 
\end{figure}

\begin{figure}[b]
\begin{center}
\begin{tabular}{c}
% trim=left bottom right top
\includegraphics[height=5.5cm, clip]{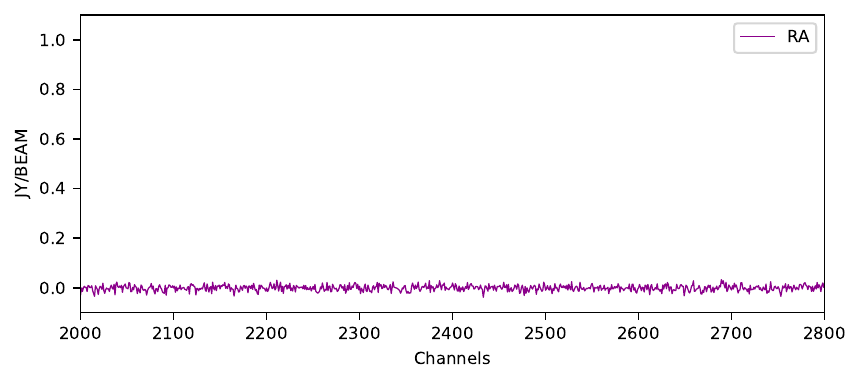}
\end{tabular}
\end{center}
\begin{center}
\begin{tabular}{c}
% trim=left bottom right top
\includegraphics[height=5.5cm, clip]{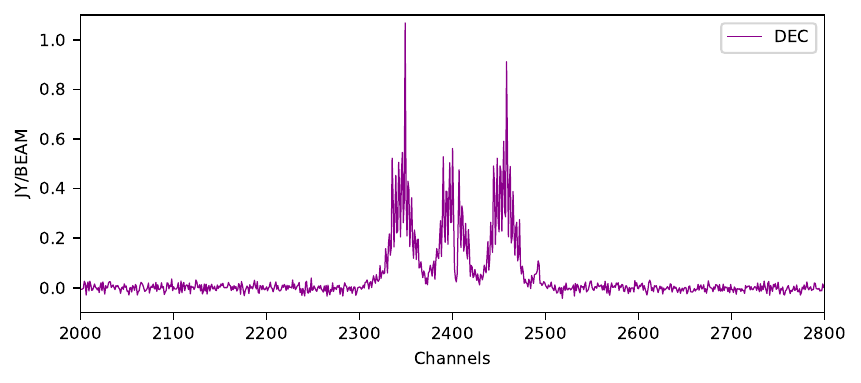}
\end{tabular}
\end{center}
\begin{center}
\begin{tabular}{c}
% trim=left bottom right top
\includegraphics[height=5.5cm, clip]{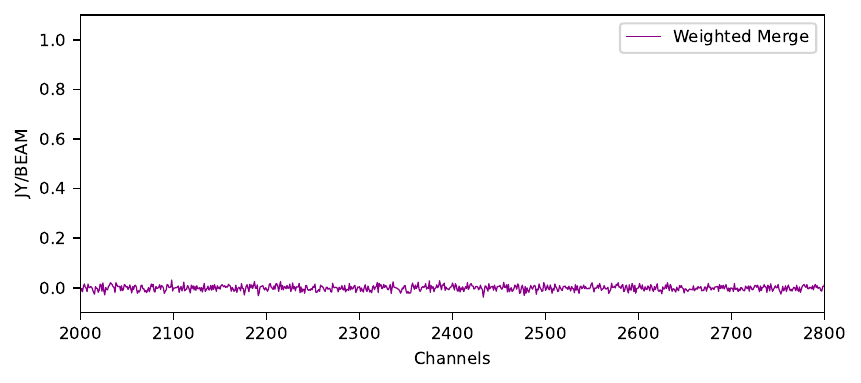}
\end{tabular}
\end{center}
\caption 
{ \label{fig:Spectra}
M31 spectra in the channel range from 2000 to 2800 extracted from a region of size $16\times16$ pixels centered in the pixel $(101,97)$. The top, middle, and bottom panels represent the RA, decl., and  weighted merge scans, respectively. The channel range corresponds to the frequency range from 6.08\,GHz to 6.16\,GHz, with channel spacing corresponding to 91.6 kHz.} 
% For conversion we use GHz=5.9+(1.5*c/16384)
\end{figure}

\begin{figure}[b]
\begin{center}
\begin{tabular}{c}
\includegraphics[height=8cm, clip]{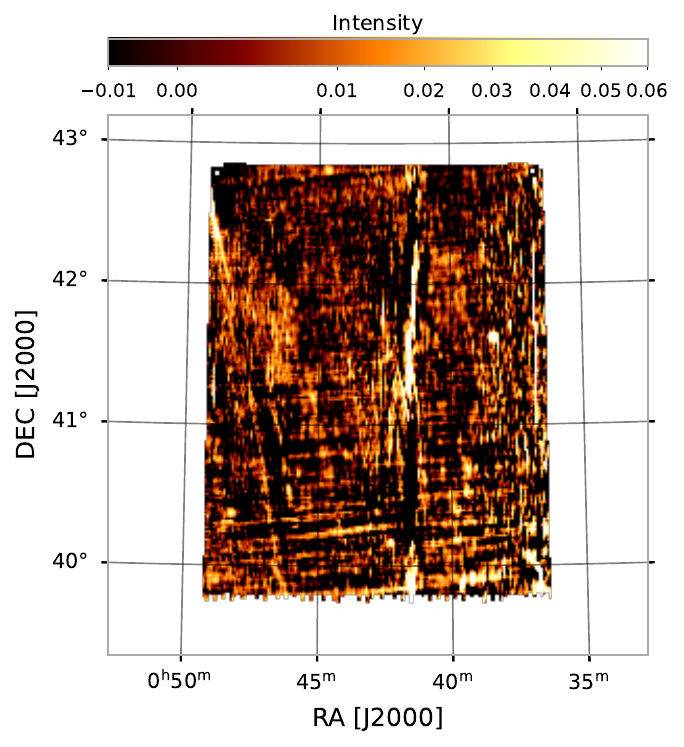}
\end{tabular}
\end{center}
\begin{center}
\begin{tabular}{c}
% trim=left bottom right top
\includegraphics[height=8cm, clip]{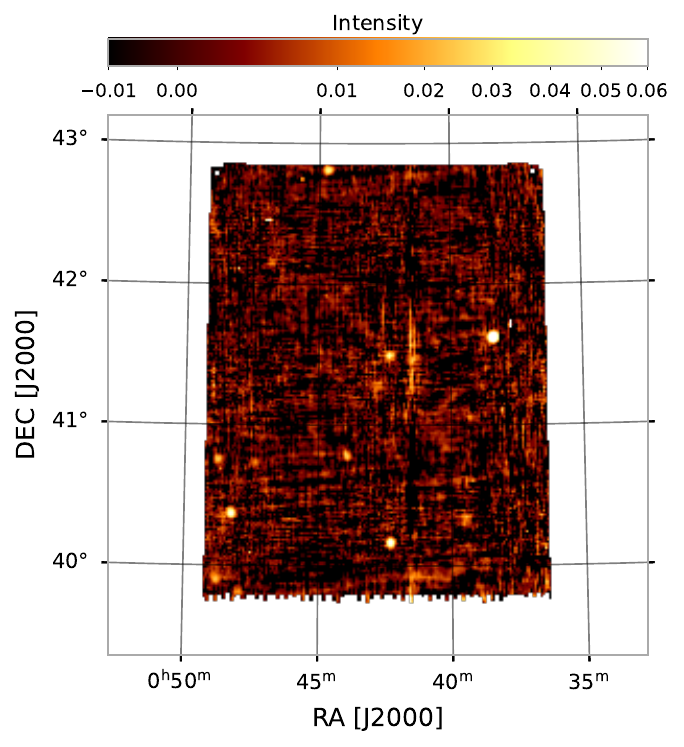}
\includegraphics[height=8cm, clip]{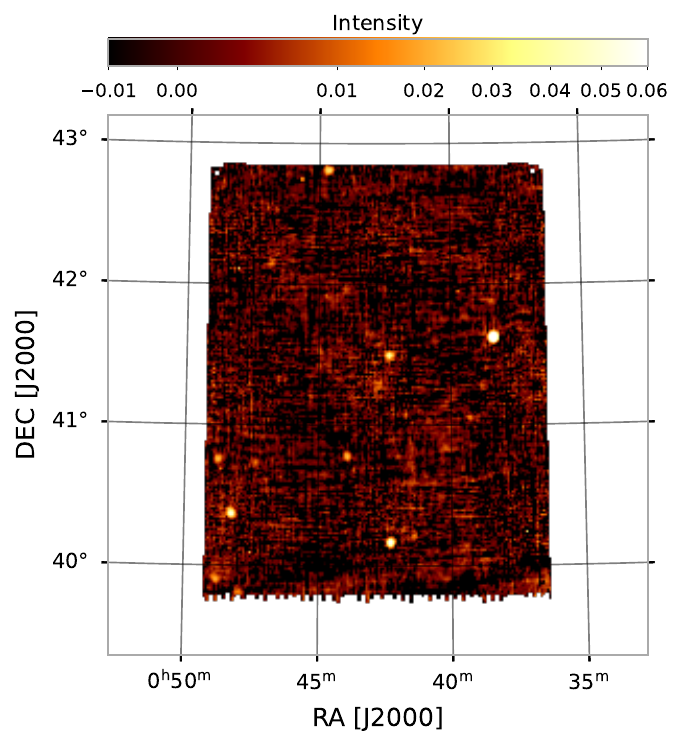}
\end{tabular}
\end{center}
\caption 
{ \label{fig:FinalResults}
M31 Images of the spectral average from frequency 6.0\,GHz to 7.2\,GHz. Top panel: direct spectral average without removing the RFIs. Bottom panels: the left panel shows the spectral average obtained after flagging the spectral channels with a sigma greater than 0.35; the right panel shows the spectral average obtained by applying the weighted stacking algorithm presented in this work.} 
\end{figure}

\subsection{Testing the Spectral-Weighting Algorithm on Real Data}

The M31 data set is composed by 22 complete scans acquired along the R.A. and decl. directions in left and right circular polarization. Each scan is projected into a spectral cube composed of 16384 channels with a channel spacing of 91.6 kHz. The observed band range is from 6.0 to 7.25 GHz (from channels 1100 to 14200), and the angular resolution is 2.9\,arcmin (FWHM beam).
For our test, we initially focus on applying the method to analyze solely two spectral cubes with a single polarization. Subsequently, we present the outcomes achieved by applying the algorithm to all 22 scans, including both left and right polarization. 
We begin by presenting the results obtained by applying the algorithm to the spectral cubes RA1 and DEC3 observed on June 29, 2016. These two cubes are representative cases of the typical RFI scenario encountered in these observations. For simplicity, we now consider only the right circular polarization.

When merging the R.A. and decl. cubes, we set the parameter $L=4$\,pixels (see Section\,\ref{Algorithm}). Consequently, for each position, we select a total of $9\times9=81$ neighboring pixels to calculate the variance and determine the weight. 
Selecting a $9\times9$ pixel region ensures a minimum of two beams on each side.
The scans are designed to be separated by 0.7 arcmin along the transverse direction. Then the data are re-projected into an image with a spatial resolution of 1 pixel=0.7 arcmin. This results in the contribution of 9 independent scans in both the R.A. and decl. to the statistics of each region.

In Fig. \ref{fig:Channel2342}, we show the specific example of channel 2342 and its respective weights for the two input cubes and the final weighted merge cube. For this channel, the R.A. scan shown on the top-left panel is free from RFIs, while the decl. scan shown on the top-right panel is heavily affected by RFIs. Therefore, this is a good case to test the \textsc{Spectral-Weighting} algorithm. In the left and right middle panels, we present the image of the local weights for the R.A. and decl. scans, respectively. The local weights of the R.A. scan are mostly uniform, except for the point at $RA=0^{h}38^{m}$ and $decl.=+41.6^\circ$ that corresponds to the brightest radio source of the field of view. The local weights of the decl. scan reveal that strong RFIs are present all over the right side. A diagonal RFI, most likely produced by a satellite, crosses the left side of the image. However, the very left part is relatively free of interfering signals. In the bottom panel, we show the final weighted merge image for this channel. The algorithm is able to successfully down-weight the RFIs while preserving the true sky emission.

To better understand the behavior of these RFIs in the nearby channels, we extract the spectra in the channel range from 2000 to 2800 from a region of size $16\times16$ pixels centered in the pixel $(101,97)$, that corresponds to the point at $RA=00^{h}54^{m}56^{s}$ and $decl.=+40.8^\circ$. This region is located on the diagonal RFI. The spectra are shown in Fig. \ref{fig:Spectra}. The top panel represents the R.A. scan spectrum, and we can observe that this region is free from RFIs. The middle panel represents the decl. scan spectrum, and it reveals that the RFI is composed of three peaks of the same intensity and width. The frequency region affected by this RFI is in the range from channels 2300 to 2500. The bottom panel represents the weighted merge scan spectrum. We observe that the RFI is completely removed from the spectrum. Furthermore, the noise in the weighted merge spectrum is appreciably smaller, as it should be after averaging the two data sets.

The right bottom panel of Fig. \ref{fig:FinalResults} shows the spectral average obtained by collapsing all the frequency channels of the weighted merge cube in one image. For reference, in the top panel of Fig. \ref{fig:FinalResults}, we show the spectral average of the merge cube obtained without applying the algorithm. Comparing the top and the right bottom images, it is clear that the RFIs are efficiently removed. Several background radio sources are clearly visible and we can even glimpse the disk of M31.

\begin{figure}[h!]
\begin{center}
\begin{tabular}{c}
% trim=left bottom right top
\includegraphics[height=8cm, clip]{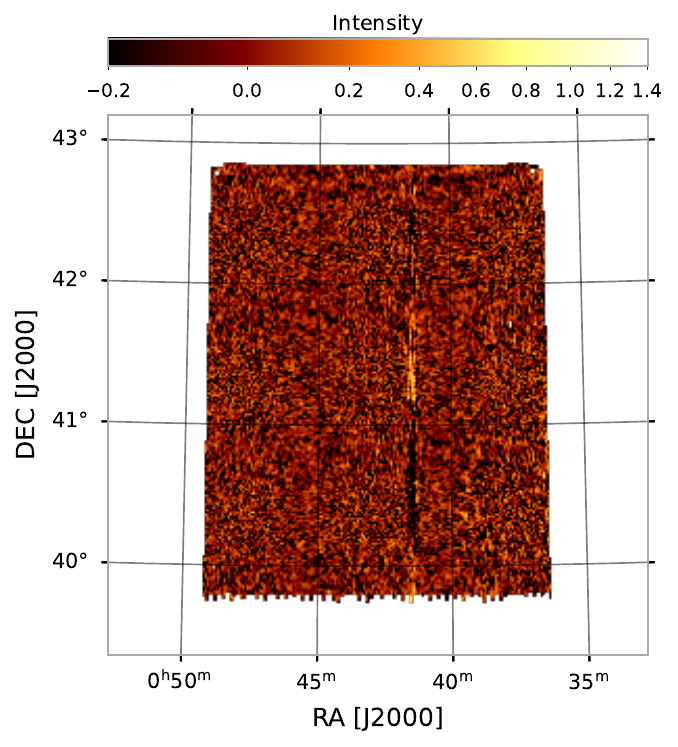}
\includegraphics[height=8cm, clip]{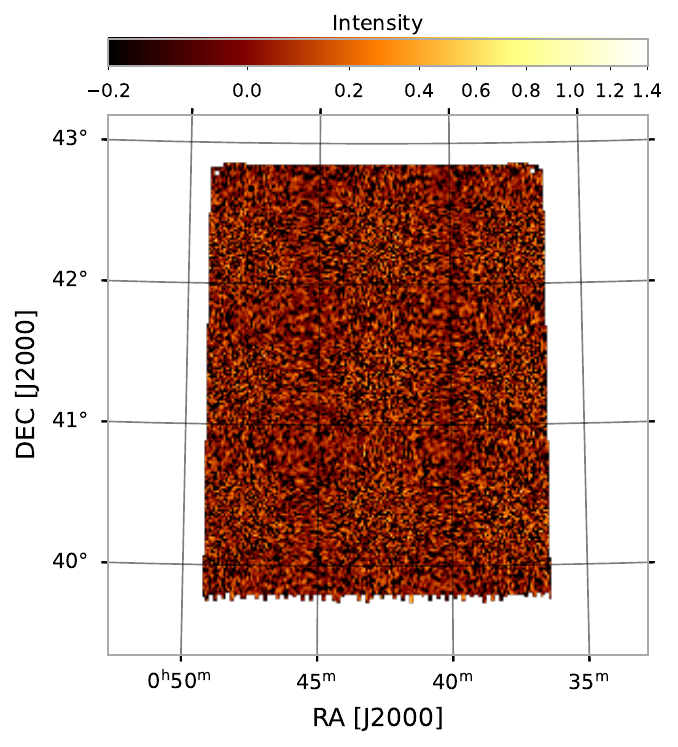}
\end{tabular}
\end{center}
\caption{ M31 images of the spectral channel 8890. Left Panel: Channel 8890 of the spectral cube obtained after flagging the input files.  Right Panel: Channel 8890 of the spectral cube obtained by applying the Spectral Weighting algorithm.} 
{\label{fig:FlagVSSpectral}
} 
\end{figure}

To compare the \textsc{Spectral-Weighting} algorithm with the flagging method, we also realize the same M31 image flagging the channels affected by interferences. First, we analyzed the spectrum of sigma noise $\sigma_N$ values of the input spectral cubes and found that $\sigma_N$ in the areas without RFIs is between 0.2 and 0.3. Therefore, we decided to set the threshold to 0.35 and flag all channels that have a sigma greater than that value. In the left bottom panel of Fig. \ref{fig:FinalResults}, we show the spectral average obtained after flagging the spectral channels with a sigma greater than 0.35. The flagging method overall reduces the RFIs, but it does not completely eliminate them. To understand why some RFIs were not effectively removed by flagging, we carefully checked the merge cube before the spectral average is performed. 
In Fig. \ref{fig:FlagVSSpectral}, we show the spectral channel 8890 of the merge cube obtained after applying both the flagging (on the left panel) and the \textsc{Spectral-Weighting} method (on the right panel). We determined that the flagging method is unable to remove the vertical RFI in this channel. The RFI is limited to a small region of the image, and thus the sigma of this channel is relatively low: $\sigma_N = 0.24 $. Since this value is lower than the threshold, the channel is not discarded, and therefore the RFI is visible in the spectral average.
On the other hand, \textsc{Spectral-Weighting} is able to completely eliminate RFI thanks to the local weights system.

\begin{figure}[h!]
\begin{center}
\begin{tabular}{c}
\includegraphics[height=8cm, clip]{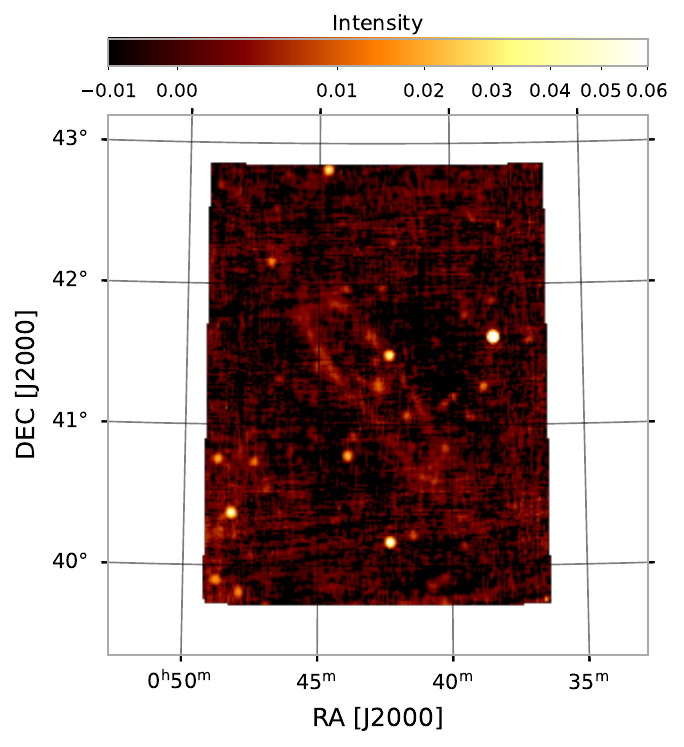}
\includegraphics[height=8cm, clip]{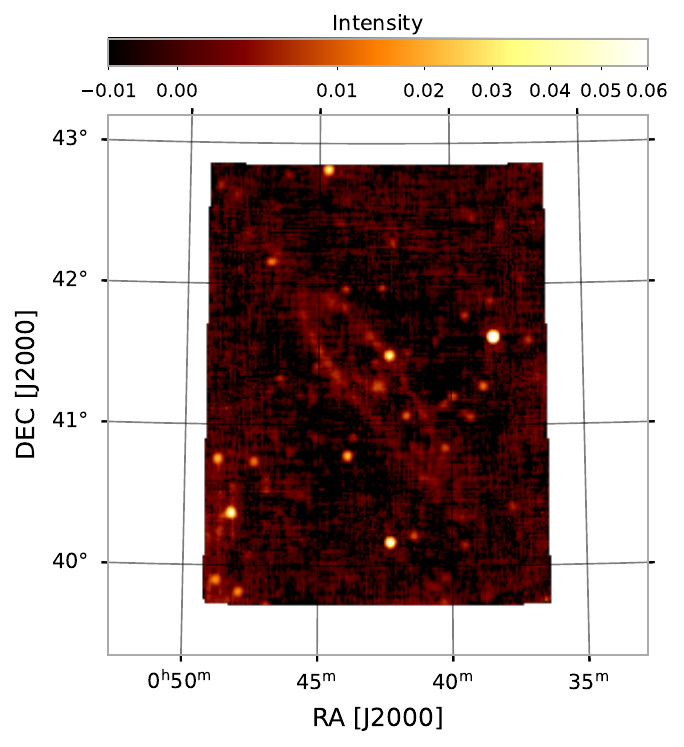}
\end{tabular}
\end{center}
\begin{center}
\begin{tabular}{c}
% trim=left bottom right top
\includegraphics[height=8cm, clip]{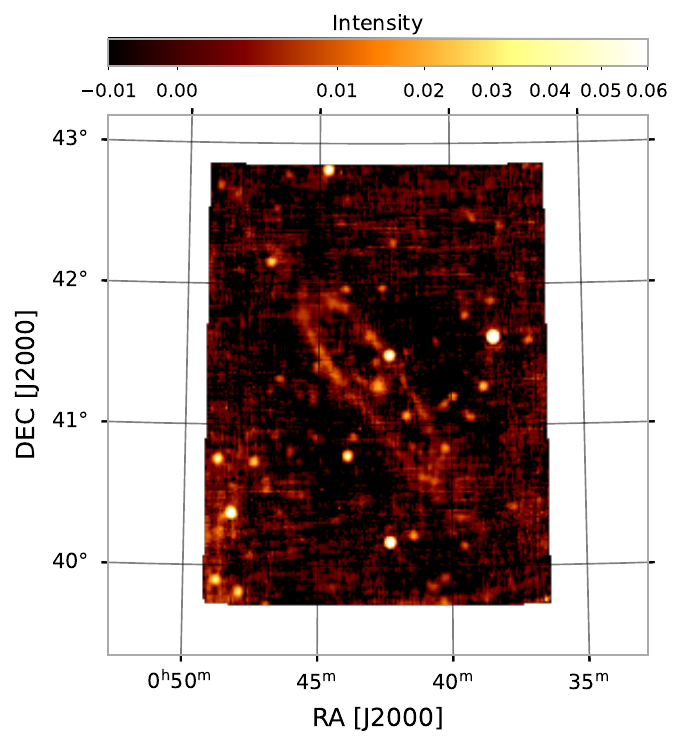}
\end{tabular}
\end{center}
\caption 
{ \label{fig:FinalResultsAllCubes}
M31 Images of the spectral average from frequency 6.0\,GHz to 7.2\,GHz obtained by combining all R.A. and decl. scans. The top-left and top-right panels respectively show the spectral average images corresponding to the left and right circular polarization spectral cubes. The bottom panel shows the total intensity image obtained by adding the left and right polarization.} 
\end{figure}

Finally, we combine all the 22 scans of the M31 data set. We apply the same procedure described above to the left and right circular polarization, separately. In the top panels of Fig. \ref{fig:FinalResultsAllCubes}, we show the spectral average images corresponding to the left and right circular polarization.
Note that RFI is generally intrinsically polarized, whereas the majority of celestial radio sources exhibit very faint circular polarization. Thus, adding together the two circular polarizations makes the process more robust. In the bottom panel of Fig. \ref{fig:FinalResultsAllCubes}, we show the total intensity image obtained by adding the left and right polarization. Due to its comparatively higher signal-to-noise, the total intensity image reveals the full extent of the M31 disk.

\section{Discussion} \label{sec:Discussion}

\subsection{Limitations and Applicability}

The \textsc{Spectral-Weighting} method is only effective for images with sources that are stable over time. It is not suitable if the sources vary over time, because the algorithm  eliminates them if a source is present in one cube but not in the others. For example, it is not appropriate to use this method when searching for variable sources, like quasars or other transient radio sources. 

We have shown that the \textsc{Spectral-Weighting} method is very effective for RFIs present in specific spectral cubes. However, a limitation arises when the same RFIs are present in all the observations because they are persistent in time. In this case, the affected channel range is ruined in all cubes and thus it is not recoverable by any means. 

In this work, we applied the \textsc{Spectral-Weighting} algorithm to single-dish images obtained using R.A. and decl. scans. However, it's essential to highlight the algorithm's versatility beyond these scans. The method applies to diverse scanning techniques, including azimuth and elevation scans, as well as spiral patterns. The fundamental requirement is the availability of repeated images of the same sky region, regardless of the scanning pattern used.
Furthermore, it's important to note that while RFIs often manifest as lines oriented along the scan direction, this is not a prerequisite for the algorithm's applicability. The algorithm effectively handles various types of RFIs, including inclined, transverse, or diffuse RFIs, as exemplified in Fig. \ref{fig:Channel2342}.

\subsection{Spectral Weighting versus Flagging}

In flagging, signals identified as RFIs are excised from the data. In the simplest application of this method, one establishes a threshold according to the expected noise level and removes the data if the signals exceed this level. However, a risk of this method is to remove a real signal like a spectral line from the data. Another problem is represented by those RFIs that are really faint and spread over a broad range of channels. These RFIs can escape the flagging threshold and appear in the spectral average. On the other hand, the \textsc{Spectral-Weighting} method does not require setting any threshold since it automatically adapts to the local noise. Indeed, the algorithm does not require any particular iteration phase to find the optimal flagging threshold. More importantly, the \textsc{Spectral-Weighting} algorithm preserves all the real signals even if they are overlapping with interfering signals, as we show in section \ref{sec:TestingOnSimulations}.

\section{Conclusions}\label{sec:Conclusions}

We implemented an algorithm based on the weighted stacking of astronomical images that combines different observations of the same region of the sky removing the interfering signals. Astronomical sources, present in all cubes, are preserved by the weighted average. However, interfering signals, present in specific cubes and a certain frequency range, are down-weighted in the average and removed from the output spectral cube.

\begin{acknowledgments}

We thank the reviewers for their valuable suggestions and feedback.
We acknowledge the hospitality of the Cagliari Observatory (OAC) where a portion of this work was performed.
The Sardinia Radio Telescope (SRT) is funded by the Ministry of University and Research (MIUR), Italian Space Agency (ASI), and the Autonomous Region of Sardinia (RAS), the European Union (EU) and is operated for the National Institute for Astrophysics (INAF) by Cagliari Observatory (OAC).

\end{acknowledgments}

\bibliography{paper}
\bibliographystyle{aasjournal}

\end{document}